\documentclass{article}

    \PassOptionsToPackage{numbers, compress}{natbib}

\usepackage[final]{neurips_2023}

\usepackage[utf8]{inputenc} %
\usepackage[T1]{fontenc}    %
\usepackage{hyperref}       %
\usepackage{url}            %
\usepackage{booktabs}       %
\usepackage{amsfonts}       %
\usepackage{nicefrac}       %
\usepackage{microtype}      %
\usepackage{xcolor}         %
\usepackage{lmodern}

\usepackage{graphicx}
\usepackage{pdfpages}
\usepackage{multirow}
\usepackage{soul}

\title{Exploring Practitioner Perspectives On Training Data Attribution Explanations}

\author{%
  Elisa Nguyen \\
  T\"{u}bingen AI Center\\
  University of T\"{u}bingen\\
  \And
  Evgenii Kortukov \\
  T\"{u}bingen AI Center\\
  University of T\"{u}bingen\\
  \And
  Jean Y. Song \\
  DGIST\\
  \And
  Seong Joon Oh\\
  T\"{u}bingen AI Center\\
  University of T\"{u}bingen\\
}

\begin{document}

\maketitle

\begin{abstract}
  Explainable AI (XAI) aims to provide insight into opaque model reasoning to humans and as such is an interdisciplinary field by nature. In this paper, we interviewed 10 practitioners to understand the possible usability of training data attribution (TDA) explanations and to explore the design space of such an approach. We confirmed that training data quality is often the most important factor for high model performance in practice and model developers mainly rely on their own experience to curate data. End-users expect explanations to enhance their interaction with the model and do not necessarily prioritise but are open to training data as a means of explanation. Within our participants, we found that TDA explanations are not well-known and therefore not used. We urge the community to focus on the utility of TDA techniques from the human-machine collaboration perspective and broaden the TDA evaluation to reflect common use cases in practice.  
\end{abstract}

\section{Introduction}
\vspace{-0.5em}
The suite of explainable AI (XAI) encompasses models and explanation methods that aim at uncovering the rationale behind black-box model behaviour for humans~\cite{guidotti2018}. XAI methods are usually attribution methods, which can be categorised into feature and instance attribution. While the former finds explanations for model predictions within the features of an input (e.g. SHAP~\cite{lundberg2017unified}), the latter explains model predictions at the instance level (e.g. Influence functions~\cite{kohliang2017}). 

This study focuses on an instance attribution approach called training data attribution (TDA). TDA gives insight by attributing model behaviour to training samples~\cite{Hammoudeh2022TrainingDI, hampel1974}.
The ground truth attribution of the model prediction on test sample $z$ to a training sample $z_j$ is the change in loss after leave-one-out retraining:
\begin{equation}
    \label{eq: loo}
    \small
    \mathrm{TDA}(z_j, z) := \mathcal{L}(z; \theta_{\setminus j}) - \mathcal{L}(z; \theta)
\end{equation}
where the model parameters $\theta$ are trained with the loss $\mathcal{L}$. As such, TDA views the model as an output of the learning algorithm and attributes model behaviour to parts of the training set. 

Explanations of machine learning (ML) models are sociotechnical in nature~\cite{ehsan2020hcxai}. Efforts in human-centred XAI emphasise this side of XAI and aim at a deeper understanding of the explainee because it is essential for the effective adoption of XAI in practice~\cite{ehsan2021explainable}. Yet, we find that the human factor of XAI is underexplored for TDA.

To address this gap, we present a qualitative interview study with ML practitioners in application areas of high-risk systems according to Annex III of the EU AI Act~\cite{aiact} (e.g. healthcare, employment, law enforcement). ML applications in such areas will require assessment throughout their product lifecycle. We therefore expect XAI to be particularly relevant in such areas. 

By interviewing practitioners, we take a human-centered perspective which gives us an impression of how ML models and explanation methods are put into practice and how practitioners view the idea of TDA. Through an inductive thematic analysis, we find: 
    (1) End-users 
    are 
    interested in training data attribution methods that could facilitate human-machine collaboration. Model developers find value in methods that enable them to improve the dataset quality. 
    (2) Though the idea of TDA is generally positively perceived, within our participant pool, TDA is not utilised. XAI tools are only used as out-of-the-box functionality. We therefore anticipate that TDA tools can deliver practical value if they are easy to implement.

\section{Related Work}
\label{sec:rw}
\vspace{-0.5em}
Interview studies provide insights into human factors in explainable AI (XAI) and can inform the design of human-centred XAI technology~\cite{liao2020questionbank}. 
Previous work has conducted semi-structured interviews with XAI practitioners of different technical expertise to study how people understand the problem of XAI~\cite{brennen2020}, people's preferences regarding interactivity of explanations~\cite{lakkaraju2022arxiv} and user needs, intentions and perceptions of explanations~\cite{kim2023help}. They found that user needs and XAI vocabulary vary across users~\cite{brennen2020} but interactivity~\cite{lakkaraju2022arxiv} and actionability~\cite{kim2023help} are desired. These studies result in concrete recommendations about XAI in practice, i.e. a call for consistent vocabulary to facilitate clear communication and progress in XAI~\cite{brennen2020}, the case for interactive dialogue systems~\cite{lakkaraju2022arxiv} and the need for considering an explanation's actionability in the design process~\cite{kim2023help}. 
However, they base their studies mainly on feature attribution explanations while our work focuses on training data attribution (TDA) explanations. We therefore expand on existing literature about user perspectives on XAI.

TDA captures a training sample's attribution to a model decision on a test sample through the counterfactual change in a model's output on a test sample when a training sample is removed from the dataset (cf. Eq.~\ref{eq: loo}). As computing TDA directly is prohibitive due to retraining costs, several methods exist~\cite{kohliang2017, guo-etal-2021-fastif, schioppa2022arnoldi, pruthi_estimating_2020, park2023trak} which focus on accurately approximating TDA. Applications of TDA methods are focused on topics from data-centric AI i.e. aiming at model improvement by improving the data (e.g. cleaning faulty data labels~\cite{teso2021interactive} or detecting model biases~\cite{pezeshkpour-etal-2022-combining, brunet2019understanding}). We find that the study of user needs and perspectives is underexplored for TDA. Our study presents a first step in addressing this gap.

\section{Interview methodology}
\label{sec: method}
\vspace{-0.5em}

\begin{figure}[t]
    \centering
    \includegraphics[width=\textwidth]{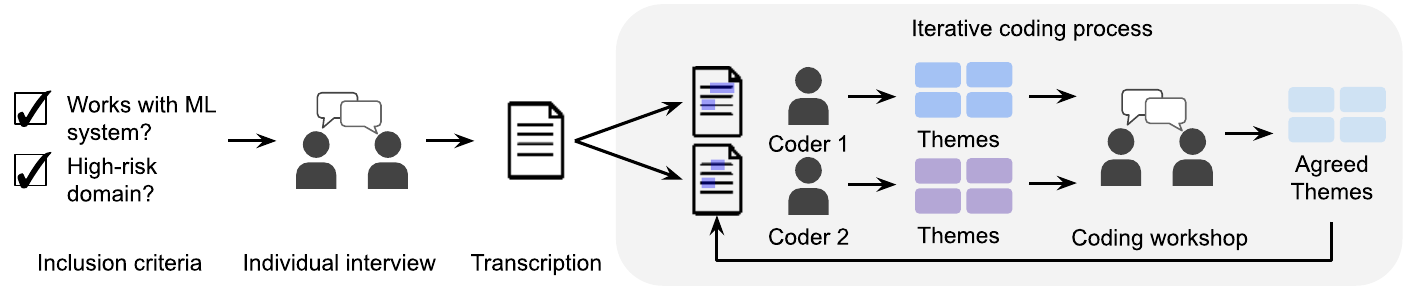}
    \caption{\small{Interview and data analysis process.}}
    \vspace{-1em}
    \label{fig:process}
\end{figure}

This study aims to explore practical perspectives on training data attribution (TDA). Since we study subjective experiences, we opt for a qualitative analysis through interviews. We conducted semi-structured interviews to balance the interview structure and the freedom of conversational flow~\cite{interviews2021} and analysed the transcripts in an inductive thematic analysis (cf. Figure\ref{fig:process}). \footnote{The IRB approval process is currently ongoing. We expect a decision in November 2023.}

\paragraph{Participants.} 
We define inclusion criteria to ensure participants align with our research aims: They should (1) have at least one month of experience in working with ML systems and (2) either work in a high-risk application area according to the EU AI Act~\cite{aiact} (e.g. health care, law enforcement, employment. Full list in Appendix~\ref{appendix:high-riskapps}). 
This criterion serves to focus our studies on 
application areas that are likely to be subject to further regulations and governance in the future~\cite{guidotti2018, doshi2017towards}.
Recruiting participants poses a challenge, especially in high-risk application areas. Hence, we use purposive sampling~\cite{guest_collecting_2023} and approach potential participants from the authors' network individually. 
We recruit 10 participants from diverse domains and degrees of experience (cf. Table~\ref{tab: participants}).

\begin{table}[t]
    \centering
    \small
    \caption{\small{Participant information. HR = Human resources, AV = autonomous vehicles, TC = telecommunications, CV = Computer vision for automation. P5 did not meet the inclusion criteria.}}
    \begin{tabular}{@{}llllll@{}}
        \toprule
        ID  & Country of work & Domain & Type & Job experience/with ML  & Type of ML\\ \midrule
        P1                & Germany         & HR  & End-user  &    3 yrs/1 mo.    & Chatbot               \\
        P2               & USA             & AV & Developer  & 2 yrs/7 yrs   & Prediction model                \\
        P3               & Netherlands     & TC & Developer  & 3 yrs/5 yrs     & Prediction model               \\
        P4                & Finland         & CV & Developer &    4 yrs/6 yrs   & Prediction model                        \\
        P6              & Switzerland     & Health    & End-user     & 2 yrs/2 yrs & Prediction model                         \\
        P7               & Netherlands     & Health & Developer      & 1 yr/3 yrs  & Prediction model                  \\
        P8             & Belgium         & Health & Developer     & 2 yrs/6 yrs     & Prediction model               \\
        P9              & Pakistan        & Health & Developer       &   5 yrs/2 yrs & Prediction model                         \\
        P10              & Germany         & HR   & End-user & 3 yrs/1 yr       & Chatbot              \\
        P11              & Germany         & Health    & End-user     &       10 yrs/6 yrs  & Clustering, Chatbot                    \\ \bottomrule
    \end{tabular}
    \vspace{-1em}
    \label{tab: participants}
\end{table}

\paragraph{Interview process.}
The interviews were conducted during June - September 2023, either in person or remotely via video call. All interviews are one-on-one conversations in English, except with P10 in German. The participants were first briefed on the objective of the study and data processing using the informed consent form (cf. Appendix~\ref{appendix:consent}). 
Upon receiving informed consent, we started the interview recording. 
Overall, the interviews lasted between 30 to 60 minutes.  
In each interview, the following topics are addressed (full interview guide in Appendix~\ref{appendix:interviewguide}):
\begin{itemize}
    \item \textbf{Job-related information.} Perspectives may vary between different domains and levels of seniority as well as experience with the ML tool. 
    \item \textbf{Interviewee's workflow with ML systems.} By asking about the workflow with the ML tool, we wish to understand the patterns of usage and challenges participants encounter. 
    \item \textbf{Perspectives on training data.} Since we investigate TDA explanations, we explicitly ask participants about 
    the role training data plays in their tasks. 
    \item \textbf{Perspectives on data-driven XAI.} 
    We address the participant's perspectives on XAI and particularly on TDA. 
\end{itemize}

\paragraph{Interview transcription.}
The interviews are first transcribed automatically using Whisper~\cite{whisper} and then cleaned up manually.
The transcript is then pseudonymised. We translated P10's German transcript to English using DeepL~\cite{DeepL}.

\paragraph{Analysis.} We analyse the transcripts through an inductive thematic analysis by two coders (cf. Figure~\ref{fig:process}). The analysis is iterative: The interview transcript of P1 is first analysed jointly in an initial coding workshop. Afterwards, coders independently code five transcripts, extending on the themes and codes found in the initial analysis.
During an intermediate coding workshop, agreements and disagreements between the coder's themes and codes are discussed. The workshop resulted in a new, merged definition of themes and codes which are used for the remaining transcripts. At the intermediate coding workshop, the interrater agreement is 77.3\% measured by the percentage of agreement participants coded to themes. 
The final coding workshop serves the same purpose - after both coders reviewed the remaining transcripts, the overlap and gaps are discussed and the final themes are agreed upon. The final interrater agreement is 80.3\%. Full analysis instructions in Appendix~\ref{appendix:analysisinstr}.

\section{Findings}
\vspace{-0.5em}
\definecolor{ourorange}{RGB}{246,178,107}
\newcommand{\ulcolor}[2][Red]{\setulcolor{#1}\ul{#2}}
\setul{-.2em}{0.4em}
\begin{figure}[t]
    \centering
    \includegraphics[width=\textwidth]{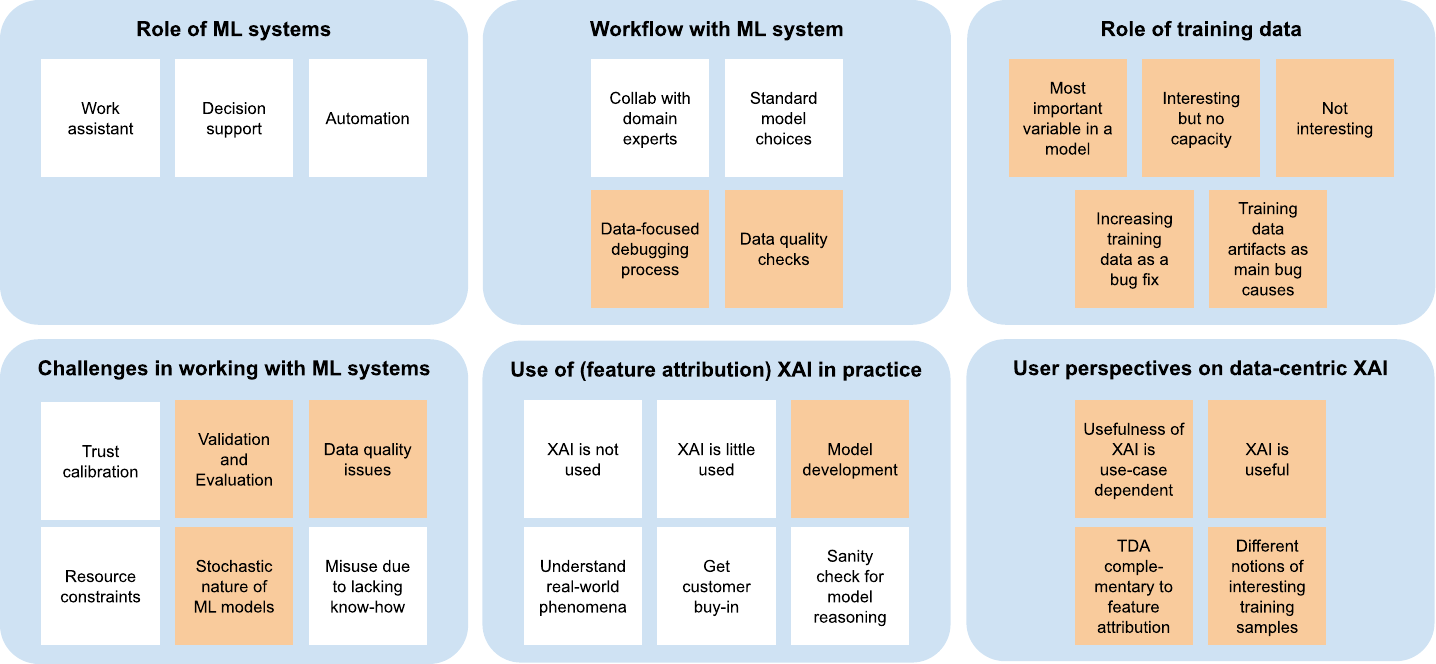}
    \vspace{-1.5em}
    \caption{\small{Theme overview as a result of the thematic coding process. Themes directly related to training data and TDA are highlighted in \ulcolor[ourorange]{orange}.}}
    \vspace{-1em}
    \label{fig:themes}
\end{figure}

The result of the thematic analysis is shown in Figure~\ref{fig:themes}. We identified six main themes which are related to the current use of ML systems, perspectives towards explainable AI (XAI) and training data attribution (TDA). Two groups of interviewees have provided noticeably different perspectives - end-users and model developers. We thus discuss their perspectives separately. 

\subsection{End-user perspective}
\vspace{-0.5em}
An end-user makes use of ML tools and is not involved in the model-building process. 
We find that end-users often face challenges related to trust calibration when using ML systems and identify a possible use of TDA in facilitating human-machine collaboration.  

\paragraph{Role of ML system.} End-users use ML systems for work assistance and decision support.
Chatbots generally fill the role of a work assistant that ``[takes work off of the participant's hands and makes their work easier]" (P10) and is ``available around the clock" (P10). Participants use chatbot systems to improve their writing in English (P1, P11), to search for information where previously they would "[ask] Google" (P1), or to ideate research ideas (P11). Moreover, P10 trusts their company-internal chatbot enough to redirect simple employee questions to the chatbot.
As decision support systems, ML systems deliver information that acts as a basis for decisions taken by end-users, e.g. diagnosis support (P6).

\paragraph{Workflow with ML system.} End-users rely on ML systems when they deliver helpful suggestions. If the ML system generates unhelpful results (P6, P10), users take over and turn away from ML support. We also find that ML systems often lack feedback loops, particularly when ML systems are purchased as a product from the market, leading end-users to voice concerns mainly when bugs accumulate (P6).

\paragraph{Role of training data.} End-user participants were unaware of which training data may have been used to train the models (P1, P10). P6 (a medical doctor) mentions that they would be curious about training data but ``[it] is a luxury that [requires time]", highlighting the practical constraint of time pressure.

\paragraph{The greatest challenge of ML system end-users is trust calibration.} Our findings agree with \citet{kim2023help}: It is unclear how much and when a system can be trusted. P1 sometimes finds themselves in a dilemma in which they wish to learn something from the chatbot, but are unable to calibrate their trust in the response due to missing knowledge: ``I don't know everything regarding this topic. I [don't even] know what he's replying to me." (P1). P10 also mentions inadequate know-how in ML system usage as a challenge: ``the employees often don't manage to ask the chatbot the right questions". 

\paragraph{Use of XAI in practice.} Not all participants have used XAI. Some participants were unaware of XAI since explanations are not a part of the ML tools they use (P1, P10, P11). P6 reports that XAI tools they used in radiology images so far (i.e. heatmaps) do not deliver a full answer to the why question, as counterfactual information is missing: ``[If] I just get an overall highlight in these basal lung regions and the prediction that is atelectasis, I still don't know why this is atelectasis but not pleural effusion or consolidation." Moreover, P6 highlights their time constraints: They would hardly be able to look at explanations even when available. Therefore, the end-user's challenges in using XAI are not only a lack of awareness and availability but also limited time.

\paragraph{Perspectives on data-centric XAI.} End-users are not familiar with the idea of TDA explanations. When asked for their opinion about the concept of TDA, chatbot users (P1, P10, P11) were interested in training samples which give additional information related to their request or samples which help them improve their interaction with the model and ask better questions (``if [the chatbot] can also sometimes formulate: [for stupid questions] you [cannot expect a good answer because] [...] then maybe you understand [how to ask the question better] and can ask it more precisely again" - P10). P6 would be interested in samples similar to the test sample to calibrate their trust. P6 also emphasised that explanations can only be helpful if there is time to spend on an explanation. Back-and-forth interactions with the system are ``absolutely unrealistic" (P6). 
The above findings agree with the insights in \citet{kim2023help}: End-users want explanations that help them improve collaboration with the ML system. End-users wish to overcome the challenge of trust calibration and showed a positive sentiment towards the idea of TDA. 

\subsection{Model developer perspectives}
\vspace{-0.5em}
Model developers are concerned with the building of ML systems. We find that model developers often face challenges related to data quality and identify potential use cases for TDA. 

\paragraph{Role of ML systems.} Model developers work on decision-support (P3, P5, P7, P8, P9) and automation systems (P2, P4). They build ML systems according to the customer's needs.
P3 uses ML systems to identify and explain the contributing factors to product issues: ``If we can predict it, we can also have an idea what are the factors mostly creating this phenomenon."  

\paragraph{Workflow with ML systems and the role of training data.} Developers and end-users collaborate closely for building and evaluating ML models, where bugs are reported to the developers by the end-users (P2, P3, P4, P9). This shows a clear separation of domain knowledge: ``Because personally, I cannot know if the model is doing the correct thing [...] business have to tell me" (P3). 
The model-building workflow is focused on data and developers spend a considerable amount of time with data curation (P2, P4, P3, P7, P8, P9). Participants explicitly stated that they use standard model architectures and the majority of the work is dataset curation (P3, P4, P8): ``[What] drives your model is your data. [...] [If] it's already an established problem, you're probably not going to do better than an algorithm that's already been laid out to solve that problem for you." (P8). Data quality checks are a set part of the data preprocessing pipeline (P2, P3, P4, P8). P2 and P9 reported that they first assess data quality before inspecting the model in debugging. Furthermore, P2 explained that collecting more data is a common way to overcome model shortcomings in autonomous driving. This shows that development work in practice is centred around data. Consequently, model developer participants consistently view training data as the most important variable in a model, e.g. ``[we] [...] believe that [...] the models can only be as good as the [...] data that you feed in." (P4).

\paragraph{Challenges in working with ML systems.} Data quality issues are often the root cause of model malfunction. Participants report distribution shifts (P3, P4), data collection artefacts like missing data or labels (P2, P4, P3, P7, P8, P9), wrong labels (P2, P4), wrong data formats due to aggregation of different data sources (e.g. dates being interpreted as integers or wrong ordering of temporal data) (P8, P9), and historical data (P9).
Issues with data quality impact model validation; for example, participants encounter difficulties due to absent labels. Furthermore, P2 mentions that the validation itself is a challenge due to multiple requirements that the ML system should fulfil. P2 also sees a challenge in the stochastic nature of ML models: ``[The] same data set, same model, you train multiple times, you can get [different] results." 
In addition, memory and compute constraints are relevant to P2 and P4 as they work with ML systems on the edge. 
Our analysis shows that data plays a substantial role both in the challenges faced by model developers and in the development process itself. 

\paragraph{Use of XAI in practice.} Participants use XAI for different purposes, most commonly as a tool for model development (P2, P3, P8). As such, XAI tools offer explanations for per-example debugging of e.g. wrong predictions 
or act as a sanity check for model reasoning.
Furthermore, P8 states that XAI tools are useful in getting customer buy-in and convincing the customers of the model's decision suggestion.
P3 described the use of XAI as a tool to understand phenomena represented by the ML model: ``[Building] the model, the whole purpose is to get some explainability. Because [...] we know that [a problem is] happening and predicting doesn't really add value. But if we can predict it, we can also have an idea what are the factors mostly creating this phenomenon." 
While XAI and therefore explanations have different purposes, we note that participants use XAI tools mainly as an out-of-the-box functionality. P3 and P8 reported using a SHAP~\cite{lundberg2017unified} library, whereas P2 visualises attention maps. We find that implementation thresholds must be low for the adoption of XAI in practice. 

\vspace{-0.5em}
\paragraph{Perspective on (data-centric) XAI.} Within our participants, we find that model developers are not familiar with TDA explanations. However, when asked about their intuition on what important training data could be, participants talked about out-of-distribution samples (P3, P8), mislabelled samples (P2), and samples close and far from the model's decision boundary (P7, P8). 
Developers seek to understand the data distribution and find ways to improve the data quality, and participants are interested in how TDA enables this. However, some participants specified that the usefulness of XAI depends on certain conditions: P3 and P8 who use explanations to present models to their business, state that in their experience, model performance must be high for explanations to serve their purpose. Additionally, P8 mentions that finding an individual training sample is unlikely to be informative in a large dataset and relevant data on a ``collection level" would be more interesting. 
Our analysis shows that the idea of TDA is positively perceived by model developers. Furthermore, TDA as a data-centric XAI approach could fit well into the work of a model developer, which is strongly centred around the data itself. 

\section{Implications for future TDA research}
\vspace{-0.5em}
\paragraph{Status quo of TDA research.} Training data attribution (TDA) explains model behaviour by finding relevant training data to a model prediction, where ``relevant" is defined by the change in loss after leave-one-out retraining (LOO) (cf. Eq.~\ref{eq: loo})~\cite{Hammoudeh2022TrainingDI, hampel1974}. As mentioned in section~\ref{sec:rw}, recent TDA research is focused on studying efficient and accurate approximations of Eq.~\ref{eq: loo} (e.g.~\cite{park2023trak}) or the application of TDA methods to particular use cases in data-centric AI (e.g.~\cite{pezeshkpour-etal-2022-combining}). The human factor in TDA is underexplored and our study takes a first step in addressing this gap. 

\vspace{-0.5em}
\paragraph{Some of the ideas from our study are actively researched.} 
Our analysis of participants' ML workflow and perspectives on XAI has shed light on the required features for TDA methods.
Some are being actively studied in the community:
P8 mentions that the attribution of a single training sample is unlikely to be informative, which has been studied in e.g.~\cite{basu_influence_2021, nguyen2023bayesian} and motivated TDA approaches like~\cite{koh2019accuracy,ilyas2022datamodels}. 
Also, model developers' intuition that mislabeled data are important training data is addressed in TDA research through existing evaluations using mislabel identification tasks as in \citet{kohliang2017}. 

\vspace{-0.5em}
\paragraph{Some are yet to be studied further.} Other perspectives could add to TDA research: Participants mention several types of data quality issues beyond mislabels, such as missing data (P3, P8, P9), wrong data formats (P8, P9), distribution shifts (P3, P4), which are currently not often considered in evaluation. Furthermore, questions related to TDA in human-machine collaboration, like interaction and usability (P1, P6, P10, P11), are not explored in TDA research. 

\vspace{-0.5em}
\paragraph{Future directions in TDA research.} It is important to consider the user and human factors in the development of XAI technology like TDA, whether it addresses model developers or end-users~\cite{ehsan2020hcxai}. We find that participants are generally unaware of TDA and therefore do not apply it even in suitable use cases. To improve accessibility, TDA researchers should understand and address user needs better. This includes, for example, expanding the current evaluation practices to cover diverse use cases. Practical constraints like time pressure (P6) and low implementation thresholds (P3, P8) should also be actively formulated as one of the research goals in the future.

\section{Conclusion}
\vspace{-0.5em}
We present a qualitative interview study with ML practitioners from various high-risk application areas to investigate the human factor of training data attribution (TDA) explanations. 
Through an inductive thematic analysis, we find that priorities and perspectives differ between end-users and developers but the idea of gaining insights into the model through training data is positively perceived overall. 
Our research reveals possible research directions in TDA to bridge the gap from research to practice: TDA for human-machine collaboration and expanding the evaluation of TDA to diverse data-centric use cases. Further, we highlight that simple and intuitive implementations of TDA methods are key.

\begin{ack}
The authors thank the International Max Planck Research School for Intelligent Systems (IMPRS-IS) for supporting Elisa Nguyen. This work was supported by the T\"ubingen AI Center.
\end{ack}

\bibliographystyle{unsrtnat}
\bibliography{references}

\begin{thebibliography}{28}
\providecommand{\natexlab}[1]{#1}
\providecommand{\url}[1]{\texttt{#1}}
\expandafter\ifx\csname urlstyle\endcsname\relax
  \providecommand{\doi}[1]{doi: #1}\else
  \providecommand{\doi}{doi: \begingroup \urlstyle{rm}\Url}\fi

\bibitem[Guidotti et~al.(2018)Guidotti, Monreale, Ruggieri, Turini, Giannotti, and Pedreschi]{guidotti2018}
Riccardo Guidotti, Anna Monreale, Salvatore Ruggieri, Franco Turini, Fosca Giannotti, and Dino Pedreschi.
\newblock A survey of methods for explaining black box models.
\newblock \emph{ACM Comput. Surv.}, 51\penalty0 (5), 2018.
\newblock ISSN 0360-0300.
\newblock \doi{10.1145/3236009}.
\newblock URL \url{https://doi.org/10.1145/3236009}.

\bibitem[Lundberg and Lee(2017)]{lundberg2017unified}
Scott~M Lundberg and Su-In Lee.
\newblock A unified approach to interpreting model predictions.
\newblock \emph{Advances in neural information processing systems}, 30, 2017.

\bibitem[Koh and Liang(2017)]{kohliang2017}
Pang~Wei Koh and Percy Liang.
\newblock Understanding black-box predictions via influence functions.
\newblock In Doina Precup and Yee~Whye Teh, editors, \emph{Proceedings of the 34th International Conference on Machine Learning}, volume~70 of \emph{Proceedings of Machine Learning Research}, pages 1885--1894. PMLR, 06--11 Aug 2017.
\newblock URL \url{https://proceedings.mlr.press/v70/koh17a.html}.

\bibitem[Hammoudeh and Lowd(2022)]{Hammoudeh2022TrainingDI}
Zayd Hammoudeh and Daniel Lowd.
\newblock Training data influence analysis and estimation: A survey.
\newblock \emph{ArXiv}, abs/2212.04612, 2022.
\newblock URL \url{https://api.semanticscholar.org/CorpusID:254535627}.

\bibitem[Hampel(1974)]{hampel1974}
Frank~R. Hampel.
\newblock The influence curve and its role in robust estimation.
\newblock \emph{Journal of the American Statistical Association}, 69\penalty0 (346):\penalty0 383--393, 1974.
\newblock \doi{10.1080/01621459.1974.10482962}.

\bibitem[Ehsan and Riedl(2020)]{ehsan2020hcxai}
Upol Ehsan and Mark~O. Riedl.
\newblock Human-centered explainable ai: Towards a reflective sociotechnical approach.
\newblock In Constantine Stephanidis, Masaaki Kurosu, Helmut Degen, and Lauren Reinerman-Jones, editors, \emph{HCI International 2020 - Late Breaking Papers: Multimodality and Intelligence}, pages 449--466, Cham, 2020. Springer International Publishing.
\newblock ISBN 978-3-030-60117-1.

\bibitem[Ehsan et~al.(2021)Ehsan, Passi, Liao, Chan, Lee, Muller, Riedl, et~al.]{ehsan2021explainable}
Upol Ehsan, Samir Passi, Q~Vera Liao, Larry Chan, I~Lee, Michael Muller, Mark~O Riedl, et~al.
\newblock The who in explainable ai: How ai background shapes perceptions of ai explanations.
\newblock \emph{arXiv preprint arXiv:2107.13509}, 2021.

\bibitem[Parliament(2023)]{aiact}
European Parliament.
\newblock Ai act, annex iii, 2023.

\bibitem[Liao et~al.(2020)Liao, Gruen, and Miller]{liao2020questionbank}
Q.~Vera Liao, Daniel Gruen, and Sarah Miller.
\newblock Questioning the ai: Informing design practices for explainable ai user experiences.
\newblock In \emph{Proceedings of the 2020 CHI Conference on Human Factors in Computing Systems}, CHI '20, page 1–15, New York, NY, USA, 2020. Association for Computing Machinery.
\newblock ISBN 9781450367080.
\newblock \doi{10.1145/3313831.3376590}.
\newblock URL \url{https://doi.org/10.1145/3313831.3376590}.

\bibitem[Brennen(2020)]{brennen2020}
Andrea Brennen.
\newblock What do people really want when they say they want "explainable ai?" we asked 60 stakeholders.
\newblock In \emph{Extended Abstracts of the 2020 CHI Conference on Human Factors in Computing Systems}, CHI EA '20, page 1–7, New York, NY, USA, 2020. Association for Computing Machinery.
\newblock ISBN 9781450368193.
\newblock \doi{10.1145/3334480.3383047}.
\newblock URL \url{https://doi.org/10.1145/3334480.3383047}.

\bibitem[{Lakkaraju} et~al.(2022){Lakkaraju}, {Slack}, {Chen}, {Tan}, and {Singh}]{lakkaraju2022arxiv}
Himabindu {Lakkaraju}, Dylan {Slack}, Yuxin {Chen}, Chenhao {Tan}, and Sameer {Singh}.
\newblock {Rethinking Explainability as a Dialogue: A Practitioner's Perspective}.
\newblock \emph{arXiv e-prints}, art. arXiv:2202.01875, February 2022.
\newblock \doi{10.48550/arXiv.2202.01875}.

\bibitem[Kim et~al.(2023)Kim, Watkins, Russakovsky, Fong, and Monroy-Hern\'{a}ndez]{kim2023help}
Sunnie S.~Y. Kim, Elizabeth~Anne Watkins, Olga Russakovsky, Ruth Fong, and Andr\'{e}s Monroy-Hern\'{a}ndez.
\newblock "help me help the ai": Understanding how explainability can support human-ai interaction.
\newblock In \emph{Proceedings of the 2023 CHI Conference on Human Factors in Computing Systems}, CHI '23, New York, NY, USA, 2023. Association for Computing Machinery.
\newblock ISBN 9781450394215.
\newblock \doi{10.1145/3544548.3581001}.
\newblock URL \url{https://doi.org/10.1145/3544548.3581001}.

\bibitem[Guo et~al.(2021)Guo, Rajani, Hase, Bansal, and Xiong]{guo-etal-2021-fastif}
Han Guo, Nazneen Rajani, Peter Hase, Mohit Bansal, and Caiming Xiong.
\newblock {F}ast{IF}: Scalable influence functions for efficient model interpretation and debugging.
\newblock In \emph{Proceedings of the 2021 Conference on Empirical Methods in Natural Language Processing}, pages 10333--10350, Online and Punta Cana, Dominican Republic, November 2021. Association for Computational Linguistics.
\newblock \doi{10.18653/v1/2021.emnlp-main.808}.
\newblock URL \url{https://aclanthology.org/2021.emnlp-main.808}.

\bibitem[Schioppa et~al.(2022)Schioppa, Zablotskaia, Vilar, and Sokolov]{schioppa2022arnoldi}
Andrea Schioppa, Polina Zablotskaia, David Vilar, and Artem Sokolov.
\newblock Scaling up influence functions.
\newblock \emph{Proceedings of the AAAI Conference on Artificial Intelligence}, 36\penalty0 (8):\penalty0 8179--8186, Jun. 2022.
\newblock \doi{10.1609/aaai.v36i8.20791}.
\newblock URL \url{https://ojs.aaai.org/index.php/AAAI/article/view/20791}.

\bibitem[Pruthi et~al.(2020)Pruthi, Liu, Kale, and Sundararajan]{pruthi_estimating_2020}
Garima Pruthi, Frederick Liu, Satyen Kale, and Mukund Sundararajan.
\newblock Estimating training data influence by tracing gradient descent.
\newblock In H.~Larochelle, M.~Ranzato, R.~Hadsell, M.F. Balcan, and H.~Lin, editors, \emph{Advances in Neural Information Processing Systems}, volume~33, pages 19920--19930. Curran Associates, Inc., 2020.
\newblock URL \url{https://proceedings.neurips.cc/paper_files/paper/2020/file/e6385d39ec9394f2f3a354d9d2b88eec-Paper.pdf}.

\bibitem[Park et~al.(2023)Park, Georgiev, Ilyas, Leclerc, and Madry]{park2023trak}
Sung~Min Park, Kristian Georgiev, Andrew Ilyas, Guillaume Leclerc, and Aleksander Madry.
\newblock Trak: Attributing model behavior at scale.
\newblock In \emph{International Conference on Machine Learning (ICML)}, 2023.

\bibitem[Teso et~al.(2021)Teso, Bontempelli, Giunchiglia, and Passerini]{teso2021interactive}
Stefano Teso, Andrea Bontempelli, Fausto Giunchiglia, and Andrea Passerini.
\newblock Interactive label cleaning with example-based explanations.
\newblock In A.~Beygelzimer, Y.~Dauphin, P.~Liang, and J.~Wortman Vaughan, editors, \emph{Advances in Neural Information Processing Systems}, 2021.
\newblock URL \url{https://openreview.net/forum?id=T6m9bNI7C__}.

\bibitem[Pezeshkpour et~al.(2022)Pezeshkpour, Jain, Singh, and Wallace]{pezeshkpour-etal-2022-combining}
Pouya Pezeshkpour, Sarthak Jain, Sameer Singh, and Byron Wallace.
\newblock Combining feature and instance attribution to detect artifacts.
\newblock In \emph{Findings of the Association for Computational Linguistics: ACL 2022}, pages 1934--1946, Dublin, Ireland, May 2022. Association for Computational Linguistics.
\newblock \doi{10.18653/v1/2022.findings-acl.153}.
\newblock URL \url{https://aclanthology.org/2022.findings-acl.153}.

\bibitem[Brunet et~al.(2019)Brunet, Alkalay-Houlihan, Anderson, and Zemel]{brunet2019understanding}
Marc-Etienne Brunet, Colleen Alkalay-Houlihan, Ashton Anderson, and Richard Zemel.
\newblock Understanding the origins of bias in word embeddings.
\newblock In \emph{International conference on machine learning}, pages 803--811. PMLR, 2019.

\bibitem[Adeoye-Olatunde and Olenik(2021)]{interviews2021}
Omolola~A. Adeoye-Olatunde and Nicole~L. Olenik.
\newblock Research and scholarly methods: Semi-structured interviews.
\newblock \emph{JACCP: JOURNAL OF THE AMERICAN COLLEGE OF CLINICAL PHARMACY}, 4\penalty0 (10):\penalty0 1358--1367, 2021.
\newblock \doi{https://doi.org/10.1002/jac5.1441}.
\newblock URL \url{https://accpjournals.onlinelibrary.wiley.com/doi/abs/10.1002/jac5.1441}.

\bibitem[Doshi-Velez and Kim(2017)]{doshi2017towards}
Finale Doshi-Velez and Been Kim.
\newblock Towards a rigorous science of interpretable machine learning.
\newblock \emph{arXiv preprint arXiv:1702.08608}, 2017.

\bibitem[Guest et~al.(2023)Guest, Namey, and Mitchell]{guest_collecting_2023}
Greg Guest, Emily~E. Namey, and Marilyn~L. Mitchell.
\newblock Collecting {Qualitative} {Data}: {A} {Field} {Manual} for {Applied} {Research}.
\newblock SAGE Publications, Ltd, 55 City Road, September 2023.
\newblock \doi{10.4135/9781506374680}.
\newblock URL \url{https://methods.sagepub.com/book/collecting-qualitative-data}.

\bibitem[Radford et~al.(2023)Radford, Kim, Xu, Brockman, Mcleavey, and Sutskever]{whisper}
Alec Radford, Jong~Wook Kim, Tao Xu, Greg Brockman, Christine Mcleavey, and Ilya Sutskever.
\newblock Robust speech recognition via large-scale weak supervision.
\newblock In Andreas Krause, Emma Brunskill, Kyunghyun Cho, Barbara Engelhardt, Sivan Sabato, and Jonathan Scarlett, editors, \emph{Proceedings of the 40th International Conference on Machine Learning}, volume 202 of \emph{Proceedings of Machine Learning Research}, pages 28492--28518. PMLR, 23--29 Jul 2023.
\newblock URL \url{https://proceedings.mlr.press/v202/radford23a.html}.

\bibitem[Dee()]{DeepL}
Deepl translator.
\newblock \url{https://www.deepl.com/translator}.

\bibitem[Basu et~al.(2021)Basu, Pope, and Feizi]{basu_influence_2021}
Samyadeep Basu, Phil Pope, and Soheil Feizi.
\newblock Influence functions in deep learning are fragile.
\newblock In \emph{International Conference on Learning Representations}, 2021.
\newblock URL \url{https://openreview.net/forum?id=xHKVVHGDOEk}.

\bibitem[Nguyen et~al.(2023)Nguyen, Seo, and Oh]{nguyen2023bayesian}
Elisa Nguyen, Minjoon Seo, and Seong~Joon Oh.
\newblock A bayesian perspective on training data attribution, 2023.

\bibitem[Koh et~al.(2019)Koh, Ang, Teo, and Liang]{koh2019accuracy}
Pang Wei~W Koh, Kai-Siang Ang, Hubert Teo, and Percy~S Liang.
\newblock On the accuracy of influence functions for measuring group effects.
\newblock \emph{Advances in neural information processing systems}, 32, 2019.

\bibitem[Ilyas et~al.(2022)Ilyas, Park, Engstrom, Leclerc, and Madry]{ilyas2022datamodels}
Andrew Ilyas, Sung~Min Park, Logan Engstrom, Guillaume Leclerc, and Aleksander Madry.
\newblock Datamodels: Predicting predictions from training data.
\newblock In \emph{Proceedings of the 39th International Conference on Machine Learning}, 2022.

\end{thebibliography}

\newpage
\appendix

\section{Inclusion criteria: High-risk application areas}
\label{appendix:high-riskapps}
We refer to the definition of high-risk application areas according to Annex III of the European Union's AI Act~\cite{aiact}. For readability, we include an overview: 
\begin{itemize}
    \item AI applications in products that require a specific level of safety: 
    \begin{itemize}
        \item Toys,
        \item Aviation,
        \item Cars,
        \item Medical devices,
        \item Lifts.
    \end{itemize}
    \item Biometric identification and categorisation of natural persons.
    \item Management and operation of critical infrastructure.
    \item Education and vocational training.
    \item Employment, worker management and access to self-employment.
    \item Access to and enjoyment of essential private services and public services and benefits.
    \item Law enforcement.
    \item Migration, asylum and border control management.
    \item Assistance in legal interpretation and application of the law.
\end{itemize}

\section{Participant information and informed consent}
\label{appendix:consent}
A censored version of the participant information and informed consent form is attached in the following: 



# Participant information

Dear participant,

We are asking you to take part in a scientific study. You will find everything you need to know about the study in this information sheet for participants.

Please read this information carefully. If you have any questions, please do not hesitate to contact us.

Around 10 - 15 people are to take part in the study.

This study is planned and conducted by XXXX XXXX (XXXX@XXXX).

Participation in the study is voluntary. If you do not wish to participate or if you later withdraw your consent, you will not suffer any disadvantages as a result.

## Why is this study being conducted?

The objective of this study is to ensure that research efforts in training data attribution methods for AI models are human-centered and focused on real use cases. These methods aim to find relevant training data that can explain the predictions and provide insights into the model's behavior. However, the theoretical definition of "relevant" may not align with practical needs.

In the field of explainable AI, researchers are actively working on providing explanations that help humans understand how the model works. Explanations come in different formats, but this study specifically focuses on relevant training data.

To gain a practical understanding of what constitutes "relevant" training data, we plan to conduct expert interviews. These interviews will involve individuals who work with or are familiar with AI systems, particularly in "high-risk" domains where transparency in the decision process is often mandated by legislation, such as the AI Act and GDPR.

## What is the study process?

For this study, you will be interviewed on your current use of AI systems and your view on training data. The interview is semi-structured, meaning that some questions are prepared but the conversation can drift off into a natural conversation. The interview will take roughly 30 minutes to one hour. The interview will be recorded and afterward transcribed. After transcription, the original voice recording will be deleted for data protection reasons.

## Would you like to learn more about the subject of the study?

If you would like to learn more about the field of explainable AI, the following material may be helpful:

- Machine Learning Explainability Workshop by Hima Lakkaraju at Stanford (https://www.youtube.com/playlist?list=PLoROMvodv4rPh6wa6PGcHH6vMG9sEIPxL)
- Interpretable Machine Learning book by Christoph Molnar, in particular, chapter 10.5 Influential Instances (https://christophm.github.io/interpretable-ml-book/influential.html)

## Who can I contact in case of questions?

If you have any further questions, please contact:

XXXX XXXX
XXXX@XXXX

## Information on data protection

In this study, XXXX XXXX (XXXX@XXXX) is responsible for data processing. The legal basis for processing is personal consent XXXX. The data will be treated confidentially at all times.

The data will be collected exclusively for the purpose of this study described above and will only be used within this framework.

The data collected also includes personal identifying data such as names and your voice.

All data by which you could be directly identified, e.g. your name or date of birth, are replaced by an identification code (pseudonymized). This makes it almost impossible for unauthorized persons to identify you.

The data will be stored at the XXXX.

We only keep the personal data for as long as it is required for the above-mentioned purpose. The data will be deleted at the latest after 10 years after the study.

For the study, we will conduct an interview with you. The interview will be recorded with a recording device. The audio recording of the interview is first stored at the XXXX. The interview will be written down verbatim (transcribed) within three months. The audio recordings are then deleted so that only the interview transcript exists. All information that would allow third parties to draw conclusions about you is changed in the transcription so that it is no longer possible to draw conclusions. The pseudonymized transcribed text is stored for 10 years in XXXX and then deleted.

We do not transfer the collected data to other institutions in XXXX.

Consent to the processing of your data is voluntary. You can revoke your consent at any time without giving reasons and without disadvantages for you. After that, no more data will be collected. The lawfulness of the processing is carried out on the basis of the consent until the revocation is not affected by this.

In the event of revocation, you can request the deletion of the collected data. The data can also be further used in the anonymized form if you agree to this at the time of your revocation.

You have the right to obtain information about the data, also in the form of a free copy. In addition, you can request the correction, blocking, restriction of processing or deletion as well as, if applicable, a transfer of the data.

In these cases, contact:

 XXXX XXXX (XXXX@XXXX)

You also have the right to complain to any data protection supervisory authority.

You can reach the supervisory authority responsible for you at:

XXXX

# Consent form

## Consent to participation

I have been informed about the study by *XXXX XXXX*. I have received and read the written information and consent form for the above study. I was informed in detail in writing and verbally about the purpose and the course of the study, the opportunities and risks of participation, and my rights and obligations. I had the opportunity to ask questions. These were answered satisfactorily and completely. In addition to the written information, the following points were discussed:

_____

_____

My consent to participate in the study is voluntary. I have the right to withdraw my consent at any time without giving reasons and without incurring any disadvantages.

**I hereby consent to participate in the above study.**

_____

Name of the participating person in block capitals

_____

Place, date Signature of **participating person**

_____

Name of the person providing the information in block capitals

_____

Place, date Signature of **informing person**

## Consent to data processing

The processing and use of personal data for the above-mentioned study will only take place as described in the information about the study.

**I hereby consent to the processing of my personal data as described.**

_____

Place, date Signature of **participating person**

_____

Place, date Signature of **informing person**



\section{Interview guide}
\label{appendix:interviewguide}

\subsection{Interview meeting process}
\begin{enumerate}
    \item Welcome the participant
    \begin{itemize}
        \item Greeting.
        \item Thank them for their time.
    \end{itemize}
    \item Briefing and informed consent.
    \begin{itemize}
        \item Jointly discuss the participant information sheet and informed consent form.
        \item Ask participants for any open questions.
        \item Sign the informed consent form.
    \end{itemize}
    \item Conduct interview.
    \item Closing.
    \begin{itemize}
        \item Ask participants for any open questions.
        \item Thank participant for their time.
    \end{itemize}
\end{enumerate}

\subsection{Interview questions}

\begin{table}[ht]
    \centering
\caption{\small{Prepared question areas and questions.}}
    \begin{tabular}{@{}p{0.25\linewidth}p{0.75\linewidth}@{}}
        \toprule
         Question topic& Intention\\
         \midrule
         \multicolumn{2}{l}{\textbf{Job-related information}}\\
         Domain of work & Different domains may have different needs.\\
         Field of work & Different countries may have different requirements for transparency.\\
         Years of expertise & The level of expertise in the task may affect the need for explanations.\\
         \midrule
         \multicolumn{2}{l}{\textbf{Understanding the current process}} \\
         Purpose of ML model & Understand the downstream task/application domain.\\
         Model functionality & Understand how the ML model assists the downstream task.\\
         Model interaction & Understand the level of interaction with the system.\\
         Evaluation of ML model & Relevant for ML developers; Understand requirements for ML models in practice and how much developers trust the evaluations.\\
         Model malfunction process & Understand the current process of dealing with unexpected model behaviour and reasons for the process.\\
         \midrule
         \multicolumn{2}{l}{\textbf{Perspectives on training data}} \\
         Type of training data & Understand the role of training data; Is the participant aware of the training data and process?\\
         Training data quality & Understand state of real-world datasets.\\         
         \midrule
         \multicolumn{2}{l}{\textbf{Perspectives on TDA}} \\
         Participant opinion on relevant training data for unexpected model output & Relevant for model developers; Understand what artefacts participants usually encounter in datasets, what is important data and why.\\
         Participant opinion on the helpfulness of knowing important data & Understand perspectives on helpfulness of data-centric information to the downstream task.\\
         Participant opinion on TDA definition & Understand perspectives on TDA. \\
         \bottomrule
    \end{tabular}
   
    \label{tab:my_label}
\end{table}

\newpage
\section{Thematic analysis instructions}
\label{appendix:analysisinstr}

\includepdf{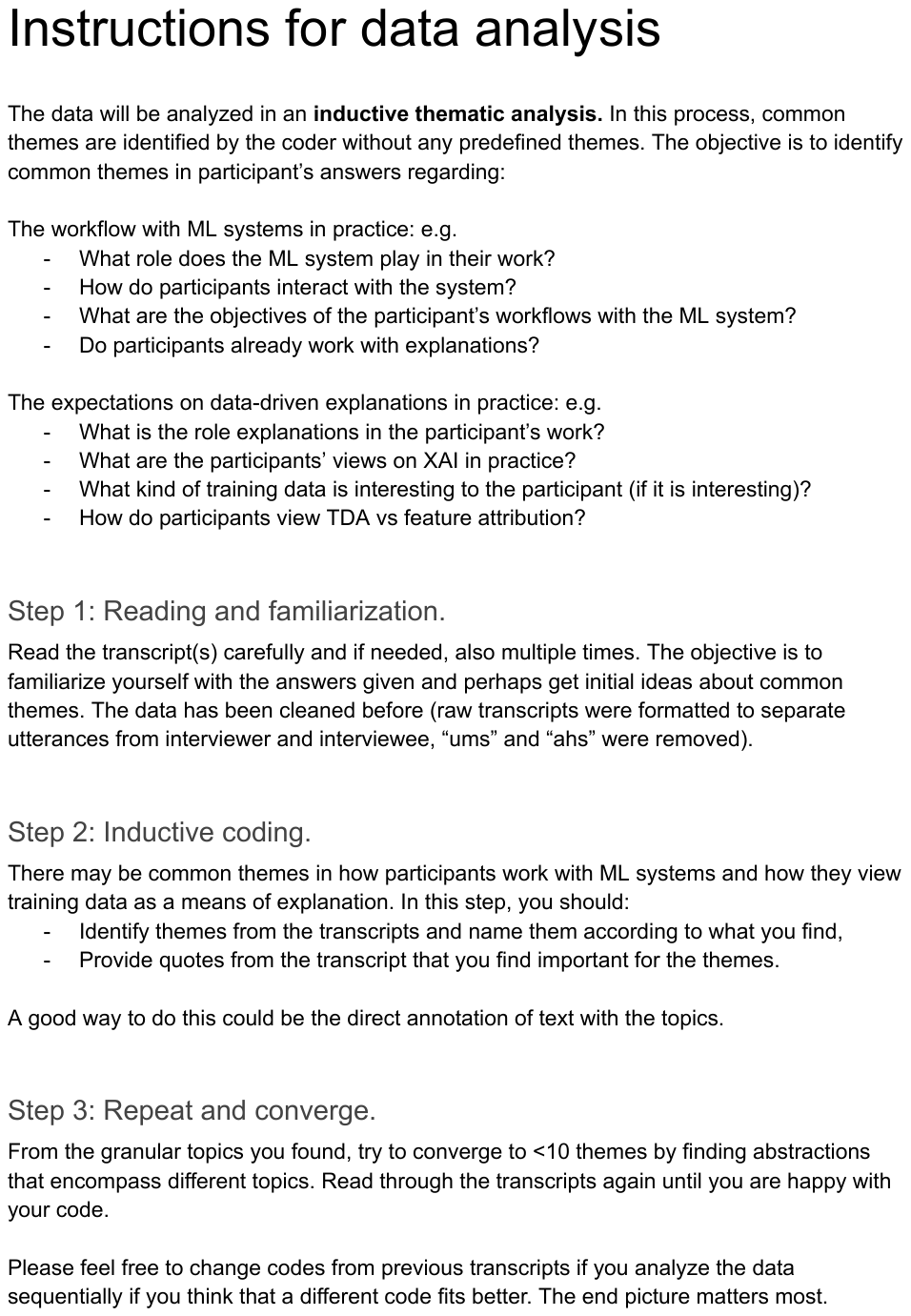}

\end{document}